\documentclass[multphys,vecphys]{svmult}

\usepackage{makeidx}         
\usepackage{graphicx}        
\usepackage{multicol}        
\usepackage[bottom]{footmisc}

\makeindex             

\begin{document}

\title*{SkyMapper and the Southern Sky Survey \\
a resource for the southern sky}
\titlerunning{SkyMapper}
\author{Stefan C. Keller, Brian P. Schmidt and Michael S. Bessell\inst{1}}
\institute{Research School of Astronomy and Astrophysics, Cotter Rd.,
  Canberra, ACT 2611, Australia
\texttt{stefan@mso.anu.edu.au}
}%
%
\maketitle

\begin{abstract}
\label{sec:1}
SkyMapper is amongst the first of a new generation of dedicated, wide-field
survey telescopes. The 1.3m SkyMapper telescope features a 5.7 square degree
field-of-view Cassegrain imager and will see first light in late 2007. The
primary goal of the facility is to conduct the Southern Sky Survey a six
colour, six epoch survey of the southern sky. The survey will provide
photometry for objects between 8th and 23rd magnitude with global photometric
accuracy of 0.03 magnitudes and astrometry to 50 mas. This will represent a
valuable scientific resource for the southern sky and in addition provide a
basis for photometric and astrometric calibration of imaging data.
\end{abstract}
\index{SkyMapper telescope}
\section{The SkyMapper Telescope}
\label{sec:2}

The SkyMapper telescope is a 1.3m telescope currently under construction by
the Australian National University's Research School of Astronomy and
Astrophysics in conjunction with Electro Optic Systems of Canberra,
Australia. The telescope will reside at Siding Spring Observatory in central
New South Wales, Australia. 

The telescope is a modified Cassegrain design with a 1.35m primary and a 0.7m
secondary. Corrector optics are of fused silica construction for maximum UV
throughput and a set of six interchangeable filters can be placed in the
optical path. The facility will operate in an automated matter
with minimal operator support. Further details on all aspects of our programme
can be found in \cite{keller}.

\section{Detectors and Filters}
\begin{figure}
\centering
\includegraphics[height=5cm]{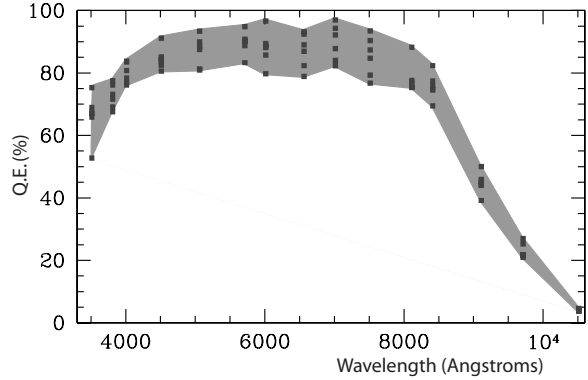}
%
%
\caption{Spectral response of SkyMapper CCDs as measured in the
  laboratory. The shaded area encloses the range in response exhibited.}
\label{fig:qe}       
\end{figure}

The focal plane is comprised of 32 2k$\times$4k CCDs from E2V, UK. Each CCD
has 2048 $\times$ 4096, 15 micron square pixels. The devices are deep
depleted, backside illuminated and 3-side buttable. They possess excellent
quantum efficiency from 350nm-950nm (see Figure\ref{fig:qe}), low read noise
and near perfect cosmetics.
%

\begin{figure}
\centering
\includegraphics[height=5.5cm]{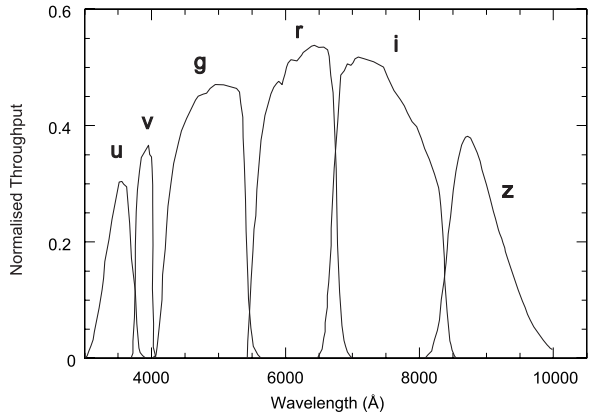}
%
%
\caption{The predicted throughput of the Southern Sky Survey filter set,
  excluding atmospheric absorption.}
\label{fig:filter}       
\end{figure}
 
The SkyMapper imager will utilise the recently developed STARGRASP controllers
developed for the Pan-STARRS project by Onaka and Tonry et al. of the
University of Hawaii \cite{onaka}. Twin 16-channel controllers enable us to
read out the array in 12 seconds with $\sim$4$-$5 electron read noise.

Figure \ref{fig:filter} shows the expected normalised throughput of our
system. The filter set is based upon the Sloan Digital Sky Survey filter set
with three important modifications:- the movement of the red edge of the $u$
filter to the blue, the blue edge of the $g$ filter to the red, and the
introduction of an intermediate band $v$ filter ( essentially a DDO38 filter
). At this time coloured glass fabrication of filters of these bandpasses
offers the best solution for spatial uniformity compared to the competing
interference film technology. Our filters are sourced from MacroOptica of
Russia.

\section{The Southern Sky Survey}
\label{sec:4}
Performing the Southern Sky Survey is the primary preoccupation of the
SkyMapper telescope. The survey will cover the 2$\pi$ steradians of the
southern hemisphere reaching $g$=23 at a signal-to-noise of 5 sigma. For stars
brighter than $g$=18 we require global accuracy of 0.03 magnitudes and
astrometry to better than 50 milli arc seconds. 

The survey's six epochs are designed to capture variability on the time scales
of days, weeks, months and years over the five year expected lifetime of the
survey. The 5 sigma limits attained after one 110 second epoch and after the
full six epochs are given in Table \ref{tab:limits}. In all bands we attain
limits slightly deeper ($\sim0.5$mag) than the Sloan Digital Sky Survey.
%
\begin{table}
\centering
\caption{Southern Sky Survey limits (5 sigma) in AB magnitudes from multiple
  110 second exposures}
\label{tab:limits}       
%
%
\begin{tabular}{lcccccc}
\hline\noalign{\smallskip}
 & $u$ & $v$ & $g$ & $r$ & $i$ & $z$  \\
\noalign{\smallskip}\hline\noalign{\smallskip}
1 epoch & 21.5 & 21.3 & 21.9 & 21.6 & 21.0 & 20.6 \\
6 epochs & 22.9 & 22.7 & 22.9 & 22.6 & 22.0 & 21.5 \\
\noalign{\smallskip}\hline
\end{tabular}
\end{table}

\section{Global Photometric Calibration}
\label{sec:5}
The greatest impediment to deriving accurate photometry from wide field
imaging cameras is the accurate description of the illumination correction. The
illumination correction corrects for geometry of the optics and inclusion
of scattered light in the system (see Patat and Freudling these proceedings). 

During commissioning we will develop an illumination correction for each
filter via dithered observations of a field. We will then rotate the
instrument and repeat the dithered observations to ensure we rigourously
understand the illumination correction for the system. We will establish six
such reference fields at declinations of around -25$^\circ$ and spaced in right
ascension. Each field will be 4.6 degrees square following the dither pattern.

During the first year of operation we will perform the Five-Second Survey, a
rapid survey in photometric conditions to provide all-sky standards between
8-16th magnitude. The Five-Second Survey will consist of a set of at least
three images of a field in all filters. 

During Five-Second observing we will observe the two highest of our six
reference fields every ninety minutes. This will ensure photometry is obtained
on a highly accurate standard instrumental system. The Five-Second Survey will
provide a network of photometric and astrometric standards to anchor the
deeper main survey images. Furthermore, it enables the main survey to proceed
in non-photometric conditions.

We will establish the six reference fields to include stars with photometry in
the Walraven system \cite{walraven}. As demonstrated by Pel \& Lub (ibid), the
Walraven system zeropoint is highly accurate: the closure solution over 2$\pi$
in right ascension has rms of less than 1 millimag. In addition, the Walraven
stars we have selected are spectrophotometric standards from the work of Gregg
et al. \cite{gregg}. The use of these standards will provide absolute flux
calibration for our system.

\section{A Filter Set for Stellar Astrophysics}
\label{sec:7}

The majority of science goals identified for SkyMapper are based on the
identification of stellar populations. It was therefore fundamental to the
science output of the telescope that we choose a filter set that offers
optimal diagnostic power for the important stellar characteristics of
effective temperature, surface gravity and metallicity. Below I will discuss
some specific examples.

Through an exploration of colour parameter space derived from model stellar
atmospheres and filter bandpasses we arrived at the filter set shown in Figure
\ref{fig:filter}. The filter set possesses two filters, $u$ and $v$,
distinctly either side of the Balmer Jump feature at 3646\AA.

\begin{figure}[ht!]
\centering
\includegraphics[height=8cm]{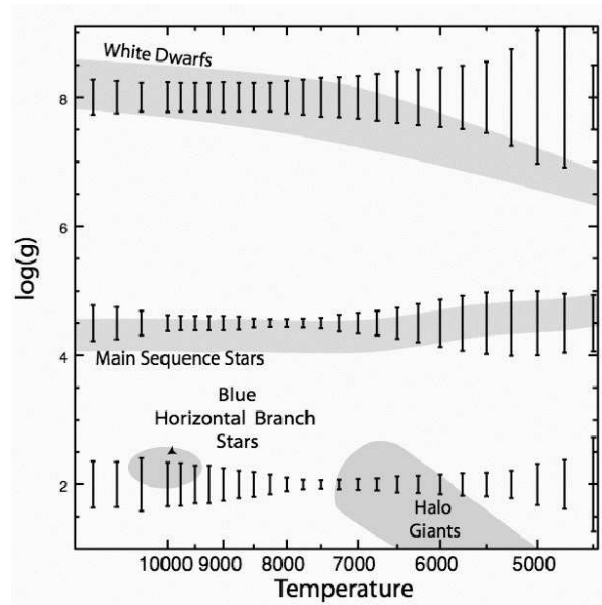}
%
%
\caption{Precision of determined surface gravity from our filter set as a
function of temperature and surface gravity (error bars show estimated
uncertainties at each point for photometric uncertainties of 0.03mag in each filter).}
\label{fig:logg}       
\end{figure}

\subsection{Blue Horizontal Branch Stars}
\index{Blue horizontal branch stars}

In Figure \ref{fig:logg} we show the uncertainty in the derived stellar
surface gravity as a function of temperature for a range of surface gravities
with photometric uncertainties of 0.03mag.\ per filter. In the case of A-type
stars we expect to determine surface gravity to $\sim10$\%. The sensitivity to
surface gravity arises from the $u$$-$$v$ colour which measure the Balmer Jump
and the effect of H$^{-}$ opacity, both of which increase with surface
gravity. It is at these temperatures that we find blue horizontal branch stars
(BHBs). Due to their characteristic absolute magnitude BHBs are standard
candles for the Galactic halo.

A line of sight through the halo inevitably contains a mixture of local
main-sequence A-type and blue straggler stars. However as is shown in Figure
\ref{fig:logg} the SkyMapper filter set enables us to clearly distinguish the
BHBs of interest on the basis of their lower surface gravity. Simulations show
that we will be able to derive a sample of BHBs to 130kpc with less than 5\% contamination.

\subsection{Extremely Metal-Poor Stars} %
\index{Extremely metal-poor stars}

\begin{figure}[hb!]
\centering
\includegraphics[height=5cm]{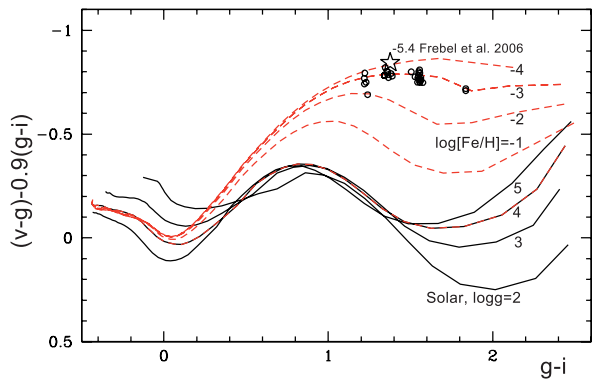}
%
%
\caption{$v$$-$$g$ vs.\ $g$$-$$i$ for stars of solar metallicity (dashed
  lines) and for a range of surface gravity (solid lines). Open circles are
  stars from the sample of \cite{cayrel} and the star symbol is HE1327-2326
  from \cite{frebel}.}
\label{fig:emp}       
\end{figure}

In the case of cooler stars (F0 and cooler) the $u$ and $v$ filters indicate
the level of metal line blanketing blueward of $\sim4000$\AA. Figure
\ref{fig:emp} shows the $v$$-$$g$, $g$$-$$i$ colour-colour diagram for a range
of metallicities and surface gravities. The $v$$-$$g$ colour has a strong
dependency on the metallicity and little dependency on the surface gravity hotter than K0 ($g$$-$$i\sim1.7$).

 This enables us to cleanly separate the
extremely metal-poor stars in the halo from the vast bulk of the halo at
[Fe/H]$<$-2. Our simulations show we should find of order 100 stars with [Fe/H]$<$-5.

\section{Data Products and Their Possible Application to ESO Calibration}

The first SkyMapper data product will be the Five-Second Survey of 8-16th
magnitude stars in the southern hemisphere. Two main survey data releases will
follow. The first data release will occur when three images in each filter
have been reduced for a field and the second (reaching 23rd magnitude in $g$)
when the full set of six have been obtained and undergone quality control.

The survey will provide sufficient density and spectral sampling of standard
stars to enable photometric calibration of any field imaged in any broadband
filter in the southern hemisphere. The largest source of dispersion in
transformations between photometric systems is due to the lack of knowledge of
the surface gravity and metallicity of the sample. The SkyMapper photometric
system provides a prior on both these points of uncertainty. Consequently we
will be able to provide improved transformations from our photometric system
to any other system. Scheduled observations on, for instance VLT or VST, may
then dispense with photometric standards and also proceed under non-photometric
conditions.

%
%
%
%
%
%
%
%

%
%



\printindex
\end{document}